\documentclass{pasj00}
\newcommand{\Hii}{H\,\emissiontype{II}~}

\usepackage{times}

\begin{document}

\title{The Hot and Dynamic Birth of Massive Stars from the ngVLA Perspective}
\author{Kei E. I. Tanaka\altaffilmark{1,2}, Yichen Zhang\altaffilmark{3}, and Kazuhito Motogi\altaffilmark{4}
}
\altaffiltext{1}{%
   Center for Astrophysics and Space Astronomy,
   University of Colorado Boulder, Boulder, CO 80309, USA: kei.tanaka@colorado.edu}
\altaffiltext{2}{%
   ALMA Project, National Astronomical Observatory of Japan, 2--21--1 Osawa, Mitaka, Tokyo 181--8588, Japan}
\altaffiltext{3}{%
   Star and Planet Formation Laboratory, RIKEN Cluster for Pioneering Research, Wako, Saitama 351--0198, Japan}
\altaffiltext{4}{%
   Graduate School of Sciences and Technology for Innovation, Yamaguchi University, Yamaguchi 753-8512, Japan}
%\email{kei.tanaka@colorado.edu}
%\KeyWords{stars: formation --- stars: massive --- binaries: close --- ISM: jets and outflows --- accretion, accretion disks --- astrochemistry}

\maketitle

\begin{abstract}
The Next Generation Very Large Array (ngVLA) has excellent capabilities to unveil
various dynamical and chemical processes in massive star formation at the unexplored innermost regions.
Based on the recent observations of ALMA/VLA as well as theoretical predictions,
we propose several intriguing topics in massive star formation from the perspective of the ngVLA.
In the disk scale of $\lesssim100{\rm\:au}$ around massive protostars,
dust grains are expected to be destructed/sublimated because
the physical conditions of temperature, shocks, and radiation are much more intense than those 
in the envelopes, which are typically observed as hot cores.
The high sensitivity and resolution of the ngVLA will enable us to detect the gaseous refractories released by dust destruction, e.g., SiO, NaCl, and AlO,
which trace disk kinematics and give new insights into the metallic elements in star-forming regions, i.e., astromineralogy.
The multi-epoch survey by the ngVLA will provide demographics of forming massive multiples with separations of $\lesssim10{\rm\:au}$ with their proper motion.
Combining with observations of refractory molecular lines and hydrogen recombination lines,
we can reproduce the three-dimensional orbital motions of massive proto-binaries.
Moreover, the 1-mas resolution of the ngVLA could possibly take the first-ever picture of the photospheric surface of an accreting protostar,
if it is bloated to the ${\rm\:au}$ scale by the high accretion rates of mass and thermal energy.
\end{abstract}

\begin{figure*}
  \begin{center}
\FigureFile(148mm,){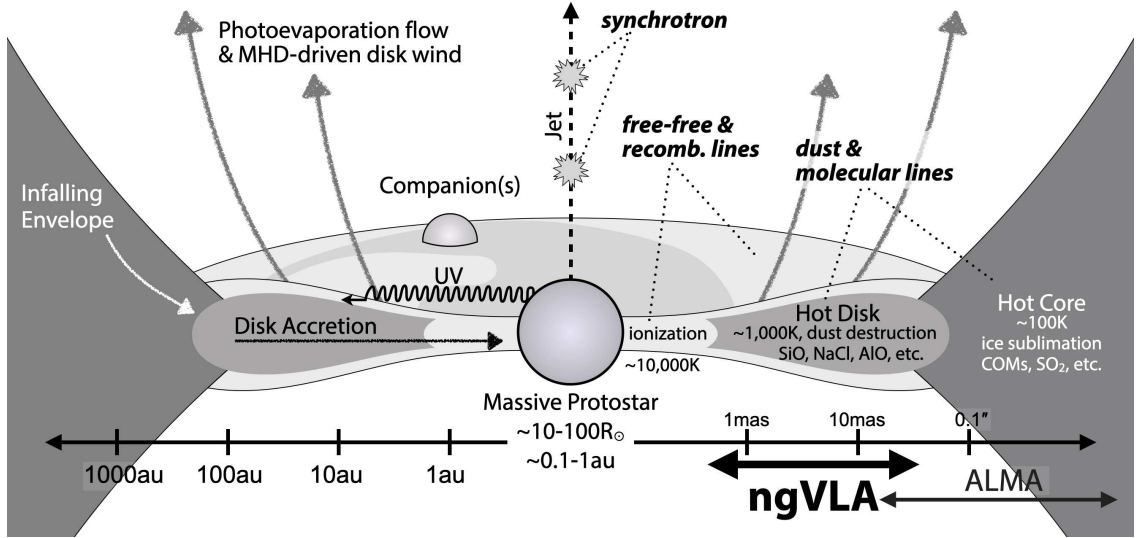} \end{center}
\caption{Schematic view of the innermost region around a massive protostar
($\gtrsim10M_\odot,~\gtrsim10^4L_\odot$).
The typical distances to massive protostars are $1$--$10{\rm\:kpc}$.
The ngVLA will be able to address various dynamical and chemical processes 
in the disk scale of $\lesssim100{\rm\:au}$,
and may directly take the first-ever picture of the photospheric surface if the massive protostar
is in the bloated phase (Sec. \ref{sec:surface}).
The three-dimensional full orbits of close companions will be recorded
in a timescale of months/years (Sec. \ref{sec:binary}).
The ``hot-disk" chemistry of refractory species such as silicon/metal compounds, 
induced by the destruction of dust grains, is an excellent topic to investigate 
by the ngVLA (Sec. \ref{sec:hotdisk}).
Multiple feedback processes, e.g., photoevaporation, MHD disk winds, and jets, 
will be spatially resolved,
and their impact on the accreting material will be quantitatively measured 
(Sec. \ref{sec:photoionization}).
Magnetic fields are not drawn to avoid confusion, but will also be traced by 
polarized emissions.
Emission mechanisms tracing these processes at radio wavelengths are 
indicated by bold italic fonts.
}
\label{fig:schematic}
\end{figure*}

\section{Introduction:\\The Innermost region in Massive Star Formation}

Massive stars ($\gtrsim10M_\odot,~\gtrsim10^4L_\odot$) are rare,
but play pivotal roles in a wide range of astrophysical and astrochemical settings in galaxies. 
However, massive star formation is still poorly understood compared to low-mass star formation,
because they are born deep in obscure clouds in large distances of $1$--$10{\rm\:kpc}$
\citep{tan14,mot18}.
The Next Generation Very Large Array (ngVLA) will uncover the hot and dynamic birth of massive stars
by its unprecedented milli-arcsecond resolution at transparent radio wavelength.

Figure \ref{fig:schematic} shows a schematic view of a massive protostar at the scale of $\sim1$--$1000{\rm\:au}$
(please note that it gives only a rough idea for reference).
Gas material accretes onto the central massive protostar from the infalling envelope
and then through the accretion disk, which is formed by the angular momentum conservation.
At the same time, multiple feedback processes eject a certain fraction of 
material from the system.
As approaching the central hot star,
the accreting material changes its physical conditions, such as temperature, chemical compositions, ionization states, and dust properties,
and thus the emission mechanism at radio wavelength changes accordingly.
The envelope and the disk are mostly neutral and traced by dust continuum and 
molecular lines;
the disk surface and the innermost disk are ionized emitting free-free continuum and 
recombination lines;
the strong shocks by the jet emit synchrotron radiation.
At present,
sub-arcsecond observations by the Atacama Large Millimeter/submillimeter Array (ALMA)
have successfully resolved the structures of infalling envelopes,
and start to report the existence of accretion disks around some massive protostars
at scales of $100$--$1000{\rm\:au}$ (see \cite{hir18, bel20}, for recent reviews).
The ngVLA will achieve an order-of-magnitude higher resolution,
addressing questions about various dynamical and chemical processes 
in the inner region of $<100{\rm\:au}$.
The ngVLA may take the first-ever picture of the protostellar surface if it is in the bloated phase.
The lower frequency of the ngVLA is also a significant advantage to look into the details of the vicinity of massive protostars,
because the dust opacity is lower and the emission lines are less crowded.
In this article,
we propose several intriguing topics in massive star formation
from the ngVLA perspective.

\section{3D Orbital Dynamics in Massive Proto-multiples} \label{sec:binary}

\begin{figure*}
  \begin{center}
\FigureFile(148mm,){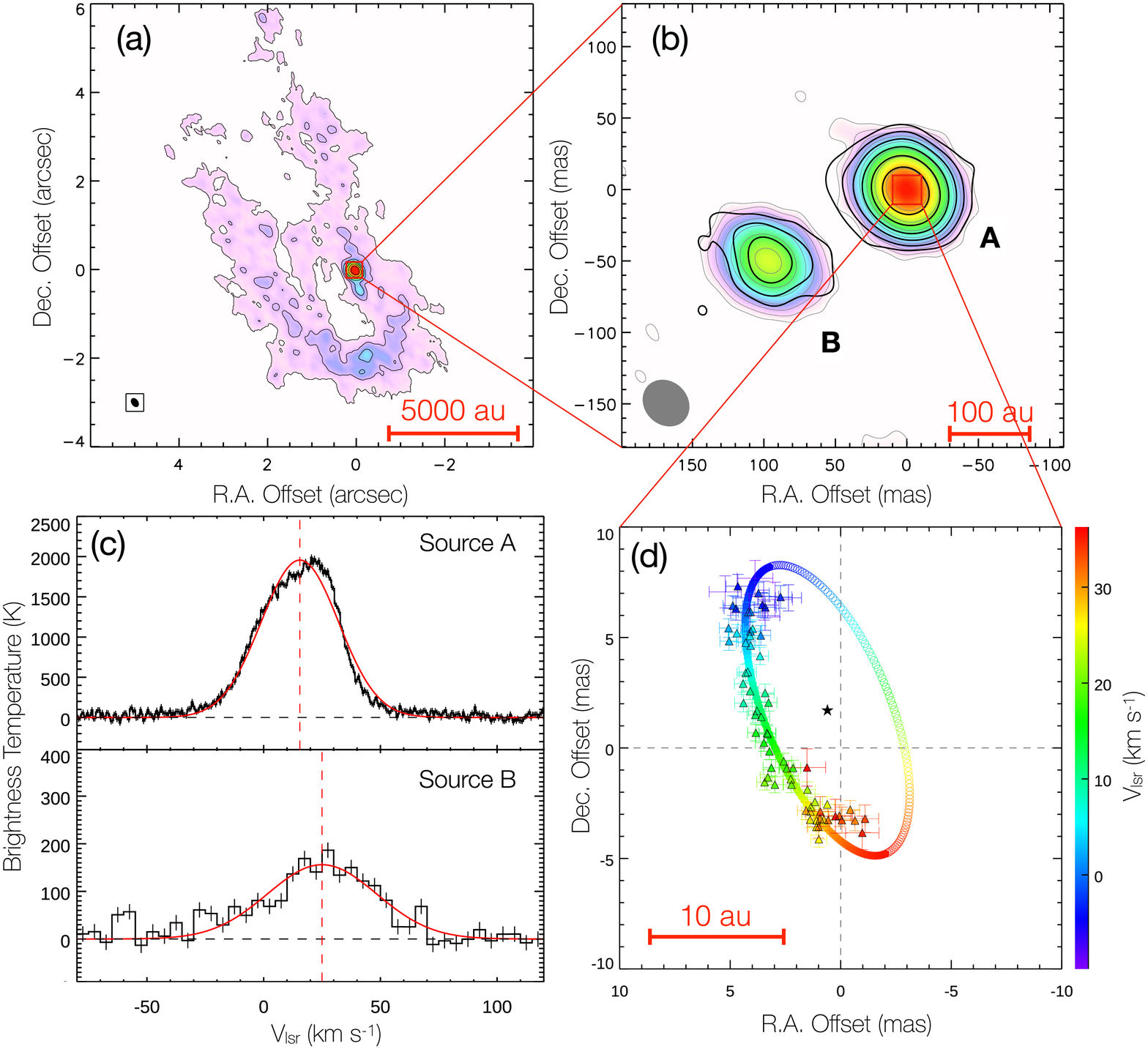} \end{center}
\caption{The massive proto-binary system in IRAS 07299--1651 \citep{zha19b}.
(a) 1.3 mm continuum on the envelope scale.
(b) The proto-binary seen by the 1.3 mm continuum (color scale and grey contours) 
and the H30$\alpha$ line emission (thick black contours).
(c) H30$\alpha$ spectra of the two protostars, showing a velocity offset of $\sim10{\rm\:km\:s^{-1}}$ due to orbital motion.
(d) Distribution of H30$\alpha$ emission centroids of source A at different
velocities, tracing the disk rotation.
}
\label{fig:binary}
\end{figure*}

The first topic is the dynamics of forming massive multiple systems.
It is well known that stellar multiplicity increases with the stellar mass,
and massive stars almost always have multiple companions, particularly at small separations \citep{duc13}.
So, massive star formation is essentially ``massive multiple formation."
Moreover, massive close binaries are important systems as the modern analog of progenitors of merging black-hole binaries.
Broadly speaking, the high fraction of close massive multiples is considered as
the consequence of disk fragmentation due to gravitational instability,
because the typical accretion rate in massive star formation is as high as $10^{-3}M_\odot{\rm\:yr}^{-1}$
(c.f., about $10^{- 6}M_\odot{\rm\:yr}^{-1}$ in low-mass star formation).
However,
the detailed dynamics of disk fragmentation and the multiplicity demographics at birth are still the least understood.
The ngVLA will dramatically improve this situation by its milli-arcsecond resolution.

The current VLA and ALMA have been reporting
massive proto-binary systems with the apparent separation of several hundred au
with their highest resolutions of $\sim20$--$50{\rm\:mas}$
(e.g., \cite{beu17}; \cite{zha19b}; \cite{KT20}; Figs. 2 and 3).
The ngVLA will discover closer systems with the separations of $\sim1$--$10{\rm\:au}$.
The significant advantage to study closer systems is the short orbital periods of
$P_{\rm orb}\sim 3000 \left(M_{\rm tot}/20M_\odot\right)^{-1/2} (a/10{\rm\:au})^{3/2}{\rm\:days}$,
where $M_{\rm tot}$ and $a$ are the total mass and the semimajor axis of the proto-binary
(for simplicity, the gravity from other companions and surrounding materials is ignored).
Therefore, massive binaries with separations of
$\sim1$--$10{\rm\:au}$ have typical orbital periods of $\sim100-3000$ days.
The ngVLA can trace the orbital motion of such massive proto-multiples on a monthly/yearly basis.
Closer companions are more common around formed massive stars \citep{san17}.
At the period of $\log P_{\rm orb}/{\rm days}\simeq3$,
formed O- and B-type main-sequence stars have the companion frequency of about $30\%$ \citep{moe17}.
Considering the later decaying of multiple systems,
the companion frequency would be even higher in the formation stage.

Continuum imaging at the higher frequencies of ngVLA,
i.e., Band 6 ($93{\rm\:GHz}$) and Band 5 ($41{\rm\:GHz}$),
is ideal for measuring the proper motions of protostars and associated substructures,
because the higher resolution can be achieved and the emission intensity is typically higher.
The continuum emission from dust and free-free would be mixed at this frequency range \citep{KT16, ros19},
but we can disentangle this degeneracy using the multi-band observations (Sec. \ref{sec:photoionization}).
While similar projects were proposed for
the studies of low-mass multiple formation \citep{tob18} and planet formation \citep{ric18},
the reward of targeting massive multiples is the high brightness temperature, as well as the high companion frequency.
The brightness temperature is estimated higher than $500{\rm\:K}$
at the scale of $\lesssim10{\rm\:au}$ around massive protostars ($\gtrsim10M_\odot,\:\gtrsim10^4L_\odot$),
which is detectable (${\rm S/N>5}$) in an hour even with the 1-mas resolution.
Therefore, the multi-epoch survey to make
the demographics of massive multiple orbits will be a low-cost, high-benefit project by the ngVLA.
We note that, even with the higher resolutions of $0.3{\rm\:mas}$,
the one-hour observation at Band 6 would be able to point out positions of massive protostars associated with the small photo-ionized regions,
because free-free emission can be as bright as $10,000{\rm\:K}$ (see also Sec. \ref{sec:protostar}).

In addition to the proper motion on the sky,
the radial motion of each protostar is measurable by emission lines.
As an example study by ALMA,
\citet{zha19b} measured the radial velocity difference of two massive protostars in IRAS 07299--1651
using H30$\alpha$ recombination line emission,
and investigated the possible binary orbits and their dynamical masses (Fig. \ref{fig:binary}).
Hydrogen recombination lines (HRLs) would be the best line tracing the orbital radial 
velocities, because they arise mostly from the vicinity of massive protostars (Fig. 
\ref{fig:schematic}).
Bands 6 and 5 cover the $n+1\rightarrow n$ transitions ($n\alpha$ transitions) with $n=38$--$44$, and $n=51$--$59$, respectively.
The velocity resolution of several ${\rm km\:s^{-1}}$ is required
to reproduce the line profile and the velocity difference of massive binaries with $a=10{\rm\:au}$.
The brightness temperature of HRLs strongly depends on the evolutionary stage of massive protostars (Sec. \ref{sec:protostar}).
If HRLs at Band 6 are as bright as $1,000$--$10,000{\rm\:K}$ (\cite{zha19a,zha19c}),
the ngVLA can achieve a significant detection (${\rm S/N>5}$) with the 5-mas resolution in 5 hours.
On the other hand, if HRLs are dimmer as $<100{\rm\:K}$,
molecular lines of refractory species, i.e., SiO, NaCl, and AlO, can be utilized instead (Sec. \ref{sec:hotdisk}).
In such a case, the 10-mas resolution with 10 hours would be required for a significant detection.
By combining the proper motion in 
continuum and the radial velocity of line emission,
we will reconstruct the three-dimensional orbital dynamics of forming massive multiples.
The observations of full orbital dynamics of massive multiples 
in early, deeply embedded phase will help us enormously
to understand the complex birth of massive multiples.

\section{``Hot Disks" with Gaseous Refractories and Highly Excited Molecules} \label{sec:hotdisk}

\begin{figure*}
  \begin{center}
\FigureFile(148mm,){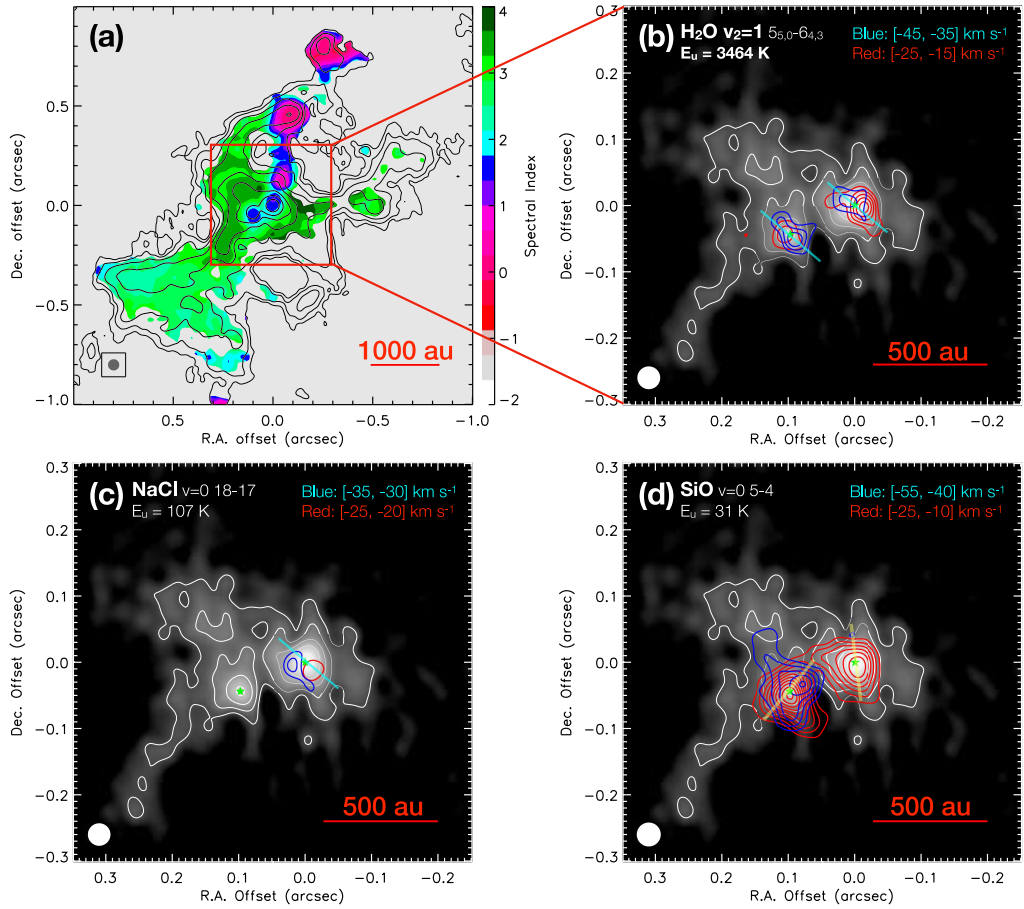} \end{center}
\caption{The massive proto-binary system in IRAS16547--4247 \citep{KT20}:
(a) The 1.3 mm continuum (black contours), and the spectral index 
($\alpha=\log{I_\nu}/\log{\nu}$) map 
between 1.3 and 3 mm (color scale).
(b--d) The integrated blue-shifted and red-shifted emission maps of three
hot-disk molecular lines, H$_2$O $(v_2=1)$, NaCl, and SiO (blue and red contours) 
overlaid on the 1.3 mm continuum emission (grey scale and white contours). 
The H$_2$O and NaCl lines tracing the rotation of the disks (with disk directions
indicated by the cyan lines), and the SiO line traces the outflow bases (with
outflow directions indicated by the yellow lines).}
\label{fig:hotdisk}
\end{figure*}

The second topic is massive protostellar disks associated with a new type of astrochemistry.
Rotating accretion disks play various essential roles in massive star formation:
close companions likely form via disk fragmentation,
magnetohydrodynamic (MHD) outflows are centrifugally driven by disk rotation,
and accretion flow circumvents super-Eddington radiation pressure to dust opacity by disk shadowing.
Detailed observations of massive protostellar disks are necessary to understand those essential dynamics in massive star formation.
ALMA recently started to discovered disks around some of massive protostars.
However, even by the highest-resolution of ALMA,
we can spatially resolve the disk structures only in limited sources
with large disks of $\gtrsim1000{\rm\:au}$ and/or close distances of $\lesssim2{\rm\:kpc}$
(\cite{mau19}; \cite{mot19}; \cite{zap19}; \cite{joh20}).
Higher-resolution observation by the ngVLA is awaited
to investigate the substructures and the dynamics inside the disks
towards statistically significant sources.

The knowledge of appropriate emission lines is the key to study the kinematics and chemical conditions.
In the envelope scale of $1000{\rm\:au}$--$0.1{\rm\:pc}$ (the outer region in Fig. \ref{fig:schematic}),
the gas temperature reaches higher than $100{\rm\:K}$,
where various complex molecules are released to the gas phase by sublimation of ice mantles of dust grains.
This phase/region of massive protostars, so-called the ``hot (molecular) cores'',
has been studied well as the cosmic laboratory of chemical processes of interstellar molecules.
The envelope dynamics are investigated utilizing the hot-core molecular lines,
such as CH$_3$CN and SO$_2$ with the energy states of $E_u/k_{\rm B}\gtrsim100{\rm\:K}$.
The astrochemical research on new heavy complex organic molecules and prebiotic molecules
in hot cores and their low-mass siblings (hot corinos) is one of the key science goals by the ngVLA
\citep{bel18a,mcg18}.

In the smaller scale of $\lesssim100{\rm\:au}$ around massive protostars (the inner region in Fig. \ref{fig:schematic}), however,
the physical conditions of temperature, density, shocks, and radiation fields are much more intense than those in hot cores.
Thus we propose different types of molecular lines as the tracers of massive protostellar disks:
(1) lines of refractory molecules, and (2) high-energy molecular lines of $E_u/k_{\rm B}>1000{\rm\:K}$,
which together we refer to as {{\it``hot-disk''} lines.
The gaseous refractory species are the products of dust destruction,
and the highly excited molecules exist due to the high temperature and strong radiation.
In the case of the massive proto-binary system IRAS 16547--4247,
\citet{KT20} detected emission lines of refractory species (NaCl, SiO, and SiS) and
vibrationally-excited water lines (H$_2$O$\:(v_2=1)$ with $E_u/k_{\rm B}>3000{\rm\:K}$)
as good probes of the individual disks (Fig. \ref{fig:hotdisk}).
These molecular lines exclusively trace the individual disks at the $100{\rm\:au}$ scale out of the $1000{\rm\:au}$ circum-binary structure.
The same types of molecular lines are also reported in disks around some other massive protostars
\citep{gin19,mau19,zha19a}.
Emission lines of aluminum monoxide (AlO), one of the most refractory materials, are detected
at the 100au scale of Orion Source I \citep{tac19}.
The presence of those hot-disk lines suggests the hot and dynamic nature of massive protostellar disks at the scale of $\lesssim100{\rm\:au}$.

These hot-disk molecular lines are excellent tracers
to investigate the kinematics of disks by the ngVLA observations.
By spatially resolving disk kinematics, the dynamical masses of the protostars will be accurately estimated.
The kinematics of substructures, such as arms, clumps, and gaps between circum-multiple and circum-stellar disks,
will directly exhibit the formation of the multiple systems, 
testing the theories of disk fragmentation.
Some hot-disk lines also trace the material at disk surfaces and above,
where the disk winds are launched and accelerated (e.g., \cite{hir17}; \cite{zha19a}; Fig. \ref{fig:hotdisk}d).
Combined with polarization observations, we can examine the MHD-driven disk wind models.
Moreover, as in the case of the orbits of multiple systems,
multi-epoch observations of hot-disk lines will enable us
to reproduce the three-dimensional dynamics of these accreting/outflowing flows.

Many transitions of refractory molecules and vibrationally excited water exist in the radio frequency of the ngVLA coverage.
For example, NaCl and KCl have their rotational transitions
from $J=1$--$0$ to $8$--$7$ and from $J=1$--$0$ to $15$--$14$, respectively
(also their vibrationally excited states).
The brightness temperature of hot-disk lines is about $50$--$150{\rm\:K}$ in IRAS 16547--4247, 
in which ALMA barely resolved the emitting spots (e.g., \cite{KT20}).
The ngVLA will spatially resolve the hot-disk regions, and the peak emission is expected to be brighter.
If the brightness temperature is $500{\rm\:K}$ at Band 5,
the ngVLA will detect the line emission (S/N$>5$) in 2 hours with the angular and velocity resolutions of 10 mas and $3{\rm\:km\:s^{-1}}$.

The radio observation of hot-disk lines is a new excellent method to investigate metallic elements in star-forming regions,
which has a unique link to meteoritics.
Primitive meteorites contains the oldest materials in our solar system, i.e., Ca-Al-rich inclusions (CAIs) and chondrules,
which were sublimated or molten once in the hot proto-solar disk.
How and where CAIs and chondrules formed is still a big mystery in planetary science.
While meteorites (and materials from sample-return spacecraft missions) keep the records in solid phase,
observations of hot-disk lines provide the complementary gas-phase information of refractory species in star-forming disks.
Further observations of the hot-disk chemistry could give unique insights into the formation of the oldest meteoritic inclusions,
and thus the origin of our Solar system.

\section{Protostellar Evolution and Photoionization} \label{sec:protostar}

One of key characteristics of massive protostars is their drastic evolution,
particularly in their radii and ionizing luminosity.
The ngVLA will let us closely observe the evolutionary stages of massive protostars,
and the dynamics of photo-ioionized gas.

\subsection{Direct Imaging of Bloated Protostars} \label{sec:surface}
Massive protostars typically accrete at the rates of $10^{-3}M_\odot{\rm\:yr}^{-1}$.
Under such high rates, protostars efficiently accumulate thermal energy and become bloated.
Theoretically,
it is proposed that the radii reach as large as $1{\rm\:au}$, or even larger, at around $10M_\odot$ \citep{hos09,hae16}.
{\it The ngVLA has the ability to directly capture the surfaces of such bloated protostars!}
If the apparent diameter of a bloated star has the apparent size larger than 1 mas (i.e., $1(d/{\rm kpc}){\rm\:au}$ in the diameter),
the ngVLA sensitivity is high enough to detect the bloated photosphere ($6000{\rm\:K}$)
in less than one hour at Bands 3--6 ($16$--$93{\rm\:GHz}$) with the 1-mas resolution.
Moreover, even with higher resolution of $\sim0.3{\rm\:mas}$,
the Band-6 observation will be able to capture the protostellar surface towards these bloated protostars
in few-hour integration time.
Fortunately, the obscuring effect by the surrounding structures would be minimal.
Dust grains should sublimate at the vicinity of the protostar.
Even if dust exists, it is likely to be optically thin at these frequencies.
Free-free by the surrounding ionized gas could be problematic at these frequencies,
but the massive protostars in bloated phase have relatively low surface temperature,
which is not enough to create \Hii regions.

\begin{figure*}
  \begin{center}
\FigureFile(148mm,){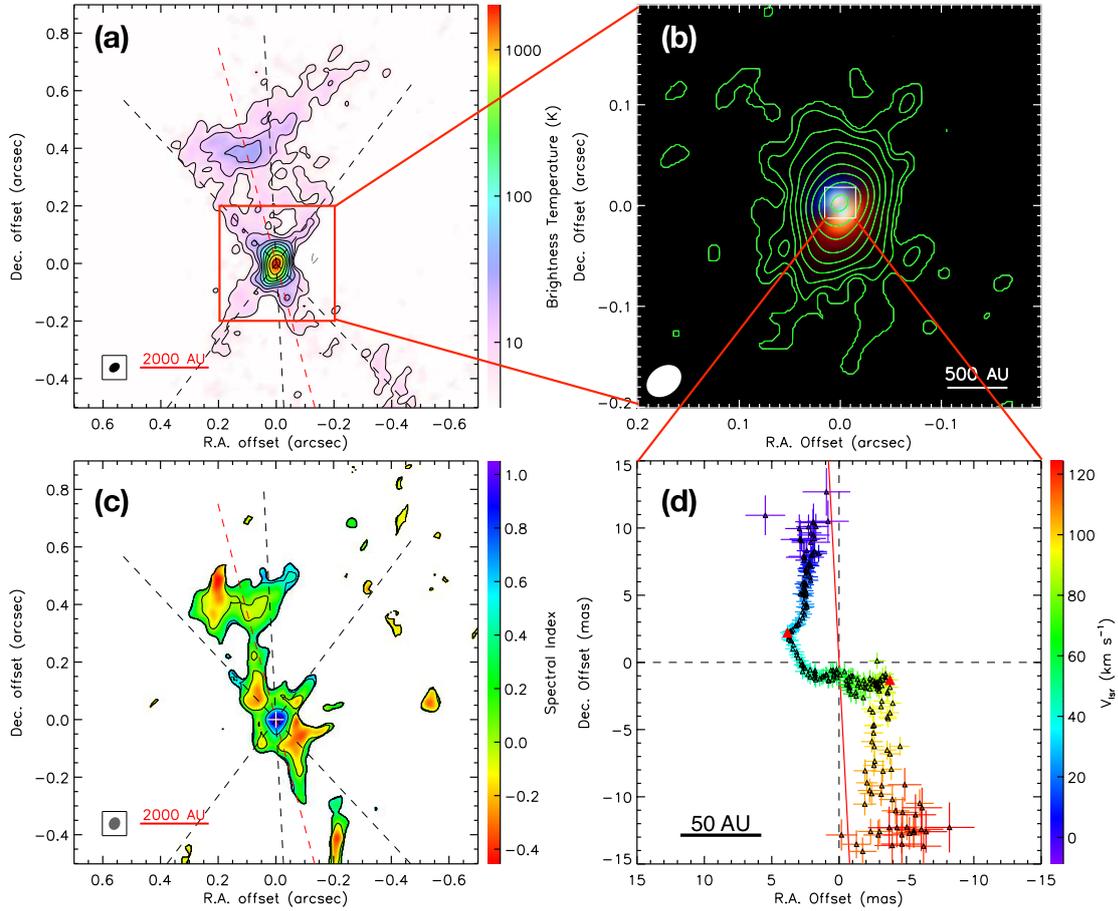} \end{center}
\caption{
The photo-evaporative outflow in the massive protostar G45.47+0.05 \citep{zha19c}:
(a) The 1.3 mm continuum emission observed by ALMA.
(b) Integrated blue-shifted and red-shifted H30$\alpha$ emissions in 
the blue and red color scales, overlaid with the
total integrated emission in green contours.
(c) The spectral index map between 1.3 and 7 mm, combining ALMA and VLA data.
(d) Distributions of H30$\alpha$ emission centroids at different velocities, 
tracing disk rotation and outflow acceleration.
}
\label{fig:G45}
\end{figure*}

Most star-formation researches have treated the central star as just a point with mass and luminosity, although it is the lead role in star formation.
The first-ever direct observations of accreting protostars will be a revolutionary leap in this research field.
The existence of massive bloated stars itself is the direct test of the theory of massive protostellar evolution,
which is also applied for the seed formation of supermassive black holes in the early universe \citep{hos13}.
Normally, stellar evolution calculations are conducted with the assumption of the spherical shape (with small oblate distortion by the rotation).
However, avoiding the spin-up to the breakup velocity,
a highly accreting star could be bar-like shape, releasing the angular momentum by the gravitational torque \citep{lin11}.
In this case, the boundary with the accretion disk would be fuzzy.
Or the magnetic fields may play important role to control the accretion rates of mass and angular momentum,
e.g., magnetospheric accretion and accretion-driven flare \citep{rom09, tak19}.
Moreover, the disk ablation by strong stellar winds could limit mass accretion at the vicinity of the protostar \citep{kee19}.
As massive stars commonly have close companions,
the au-scale radii of bloated stars would enhance event rates of the stellar interaction, such as mass transfer or merger,
that may cause the explosive outcomes as seen in some planetary nebulae \citep{sok12}.
The dynamical timescale at the stellar surface is shorter than a year,
and thus those dynamical interactions of the star and the disk can be recorded like ``movies" by the ngVLA.

\subsection{Photoionization Traced by Free-Free and HRLs} \label{sec:photoionization}

After the bloated phase, as the stellar mass grows to $10$--$20M_\odot$,
the accreting protostar turns into the Kelvin--Helmholz contraction approaching to the zero-age main sequence (ZAMS) star.
The ionizing photon rate rapidly explodes as the contraction proceeds,
because the peak of the stellar spectrum goes beyond the Lyman limit of $91.2{\rm\:nm}$.
A photoionized region is formed around the massive protostar, and the radio luminosity increases by orders of magnitude.
The stellar surface observation becomes difficult in this phase.
Instead, the ionized gas provides observational diagnostics of the protostellar evolution and the photoionization feedback \citep{KT16}.
As an example, \citet{zha19c} discovered a large ($>1000{\rm\:au}$) photo-evaporative outflow from the very massive protostar G45.45+0.05,
using high-resolution observations of ALMA at 1.3 mm and VLA at 7 mm (Fig. \ref{fig:G45}).
The modeling of free-free emission provided the density and temperature profiles of the photoionized gas.
Also, ${\rm H}30\alpha$ line emission traces the outflowing kinematics of the photo-evaporation flow.
Since the central protostar of G45 is already evolved and very massive as $40M_\odot$,
its photo-evaporation flow is large and bright enough to be investigated in detailed by current ALMA and VLA.
The ngVLA will be able to conduct similar studies towards less-evolved and dimmer sources (see also \cite{gal18}),
and test the feedback models of photo-evaporation and MHD-driven outflows/jets \citep{KT17,kui18}.

The radio continuum from massive protostars contains three components of
dust, free-free, and synchrotron (Fig. \ref{fig:schematic}).
We can solve the degeneracy of these emissions by the multi-band observations,
since each component typically has different spectral index, i.e., $\alpha=d \log{I_\nu}/ d \log{\nu}$:
$\alpha\gtrsim2$ for dust, $\alpha\simeq0$--$2$ for free-free, and $\alpha\lesssim0$ for synchrotron, respectively.
Figure \ref{fig:G45}c shows the index map between 1.3 and 7 mm of G45.45+0.05,
that unveils the collimated jets ($\alpha\lesssim0$) surrounded by the bipolar photo-evaporation flow ($\alpha\sim0$--$1$).
The presence of the synchrotron jets indicates that the disk accretion is still ongoing in spite of the strong photo-evaporation feedback.
In the case of the massive proto-binary system IRAS16547--4247 (Fig. \ref{fig:hotdisk}a),
the index map nicely illustrates
the dusty circum-binary structure ($\alpha>2$), the twin protostars with ionized gas ($\alpha\sim1$), and the aligned jet knots ($\alpha\lesssim0$).
By separating these components,
we are able to estimate the dust masses, the ionizing luminosities of the protostars, and possibly the magnetic field strength at the jets.
The multi-band imaging by the ngVLA with its high sensitivity and resolution will let us to seek into
very first stage of the creation of the \Hii regions, as well as the radio jets,
in details towards the significant numbers of sources.

HRLs are good tracer to investigate the kinematics of the ionized gas.
Despite that many transitions of HRLs are observable 
in the frequency range of the ngVLA, they are typically brighter 
at higher frequencies. Meanwhile, the free-free emission
is also optically thinner in higher frequency bands.
Thus, the HRL observations at Bands 5 and 6 will be important to investigate the innermost kinematics (see also Sec. \ref{sec:binary}).
In the case of G45.45+0.05,
the ${\rm H}30\alpha$ emission ($231.90093{\rm\:GHz}$) is strong enough
so that we are able to probe the small shift of its emission centroid
over velocities (Fig. \ref{fig:G45}d). Such shift of emission centroids
trace the disk rotation and the outflow acceleration in the $100{\rm\:au}$ scale,
which is much smaller than the beam size.
Similarly, Figure \ref{fig:binary}d shows the ${\rm H}30\alpha$ centroids in IRAS 07299--1651,
probing the rotation of the ionized disk at the $10{\rm\:au}$ scale.
Band 5 and 6 HRLs can trace the kinematics of the innermost ionized-gas 
in a similar way, and the ngVLA may even resolve such inner regions without centroid analysis.
Theoretical studies proposed that
the radiative MHD processes at the innermost scale of $1$--$100{\rm\:}$
determine the total impact of feedback toward the core scale of $0.1{\rm\:pc}$:
the disk shadowing allows the mass accretion under the super-Eddington luminosity,
the UV absorption at the dust sublimation front significantly regulates the mass loss by photo-evaporation,
and the MHD jets/outflows are driven by disk accretion/rotation
(e.g., \cite{mat17,KT17,kui18,ros20}).
The high-sensitivity and resolution of the ngVLA with the bright HRLs will
enable us to directly observe the central engines of those feedback processes.
At lower frequencies, HRLs become less bright and the free-free emission from the innermost region is likely to be optically thick.
However, with stacking, HRLs at lower frequencies could still be useful to trace the photo-evaporative flow in larger scales.

\section{Summary and Discussion}

We discussed potential research topics in massive star formation from the ngVLA perspective.
The superb resolution of the ngVLA will allow us
to investigate the unexplored inner regions of $1$--$100{\rm\:au}$ around massive protostars.
Various structures on disks, i.e., arms, gaps, disk winds, as well as companions, will be captured in detailed.
Moreover,
the ngVLA would be able to take the first-ever picture of the surfaces of the accreting protostars, if they are in the bloated phase.

Thanks to the short dynamical timescale and the bright line emissions of the innermost regions,
we will be able to reproduce the three-dimensional dynamical motions, such as companion migration,
disk fragmentation, and accreting/outflowing flows by the multi-epoch observations.
Particularly, the demographics of forming massive binary orbits is quite important because formed massive stars almost always have close companions.

While heavy complex organic molecules and prebiotic molecules are the key targets of the ngVLA astrochemistry,
these molecules are not suitable for exclusively tracing the innermost region of massive star formation.
Instead, we propose the new sets of emission lines as the tracers of the hot, dynamic disks at the scale of $<100{\rm\:au}$: (1) lines of refractory molecules, e.g., silicon compounds and alkali metal halides, (2) high-energy molecular lines of $E_u/k>1000{\rm\:K}$, e.g., vibrationally excited water,
and (3) hydrogen recombination lines from ionized gas of $10,000{\rm\:K}$.
The radio observation of the gaseous refractory species will also provide new insights into the metallic elements in star-forming regions, which has a unique connection to astromineralogy and meteoritics.

We focused on the smallest scale at $1$--$100{\rm\:au}$ in this article.
However, we cannot make the comprehensive scenario of massive star formation without understanding the larger-scale processes,
which dominates the total mass budget.
The high-sensitivity of the ngVLA will also very useful to investigate the larger-scale dynamics,
e.g., core-scale fragmentation, disk formation, mass-loading processes by large-scale outflows (e.g., \cite{bel18b}).
Toward the new era of ngVLA,
we also need conduct the detailed theoretical/numerical studies of $<100{\rm\:au}$, e.g.,
driving MHD/radiative disk winds,  dust growth/destruction, hot-disk chemistry, and the star-disk interaction.
Particularly, synthetic observational modeling with the ngVLA capabilities based on those studies will
maximize the potential of the ngVLA and help us to understand the dynamics and chemistry in massive star formation.

\end{document}